\documentclass{ws-ijmpa}

\usepackage{graphicx}% Include figure files

\newcommand{\beq}{\begin{equation}}
\newcommand{\eeq}{\end{equation}}
\newcommand{\beqa}{\begin{eqnarray}}
\newcommand{\eeqa}{\end{eqnarray}}
\newcommand{\bd}[1]{ \mbox{\boldmath $#1$}}

\begin{document}
\def\ii{\'\i}

\markboth{Peter O. Hess and Walter Greiner}
{PSEUDO-COMPLEX GENERAL RELATIVITY}

\title{
PSEUDO-COMPLEX GENERAL RELATIVITY
}

\author{Peter O. Hess$^{1}$ and Walter Greiner$^{2}$}

\address{
$^1$Instituto de Ciencias Nucleares, UNAM, Circuito Exterior, C.U., A.P. 70-543,
04510 M\'exico D.F., Mexico \\
$^2$Frankfurt Institute of Advanced Studies, Johann Wolfgang Goethe Universit\"at,
Ruth-Moufang-Str. 1, 60438 Frankfurt am Main, Germany \\
}

\maketitle

\begin{history}
\received{Day Month Year}
\revised{Day Month Year}
\end{history}

\begin{abstract}
An extension of the theory of General Relativity is proposed, based
on pseudo-complex space-time coordinates.
The new theory
corresponds to the introduction of two, in general different, metrics which are
connected through specific conditions. A pseudo-complex Schwarzschild solution
is constructed, which does not suffer any more by a singularity.
The solution indicates a minimal radius for a heavy mass object.
Consequences
for the redshift and possible signatures for its observation are discussed.

\end{abstract}
\ccode{PACS numbers: 02.40.ky, 98.80.-k}

\maketitle

\section{Introduction}

Several attempts have been made to generalize the Theory of General
Relativity (GR) by extending algebraically the real coordinates of
space-time to larger domains. For example, in Refs.
\cite{einstein1,einstein2} complex
coordinates were proposed in an attempt
to unify GR with electro-magnetism.
Recently, in \cite{mantz}, a more profound description with complex
variables has been given.
In Refs.
\cite{crumeyrolle1,crumeyrolle2,clerc1,clerc2} the coordinates were
instead extended to so-called hypercomplex variables (an equivalent notation
for pseudo-complex). There are many others, like
\cite{caneloni,brandt1,brandt2,brandt3,beil1,beil2,beil3,beil4,moffat1,moffat2,kunstatter},
which in addition exploit Born's
principle of complementary \cite{born1,born2}.
They have in common that a maximal acceleration appears in the theory,
corresponding to a minimal length scale $l$.

In \cite{cond1} it was shown that not all algebraic extensions of the
coordinates make sense. Only real and
pseudo-complex coordinates allow for
physical solutions, i.e. the non-appearance of ghost solutions.
This finding suggests to study in more detail the algebraic extension
of the space-time coordinates to pseudo-complex variables,
also investigating alternative proposals.

In \cite{paper1,paper2} a pseudo-complex extension of Field Theory
was presented, with the intention to investigate a modified dispersion
relation, which allows a shift of the GZK limit \cite{gzk1,gzk2}
in the spectrum of high energy cosmic rays.
Quite interesting, this theory is automatically regularized! Deviations
of some cross sections, with respect to standard results, were calculated
resulting in a theoretically deduced upper limit of the minimal length scale $l$ of approximately
$10^{-28}$~cm. The approach presented in \cite{paper1,paper2},
partially relies on proposals made in \cite{schuller0,schuller1}.
The theory does conserve Lorentz invariance, which is attractive.
Usually, the appearance of a minimal length is identified with the violation
of Lorentz invariance. In contrast, in \cite{paper1,paper2} the minimal
length scale appears as a {\it scalar parameter} in the theory, not affected
by Lorentz contraction.

The fact that pseudo-complex variables seem to be useful in distinct
areas of physics, justifies again a deeper investigation.
Our motivation is to study the algebraic extension
of the space-time coordinates to pseudo-complex variables. The questions
posed are: Is there a possibility to formulate this extension in a consistent
manner? Because in field theory the appearance of a minimal length scale
renders it regularized, hence: does the same happen in GR? Do singular solutions
render non-singular? What is the role of the minimal length scale? What is
the structure of the equivalent Schwarzschild solution, i.e., of a
spherical symmetric large mass? Do black holes become gray?

We can not answer all these questions in this contribution and refer
to a later publication. Nevertheless, some questions will be addressed:
We will be able to formulate in a consistent way the extension of GR to pseudo-complex numbers. A solution for a spherical
symmetric mass distribution, equivalent to the standard Schwarzschild
solution, will be presented.

We proceed as follows:
Based on the Refs. \cite{paper1,paper2},
we will extend GR to a pseudo-complex description
of the coordinates, using a different approach as in
\cite{crumeyrolle1,crumeyrolle2,clerc1,clerc2}.
It will include a new variational principle, as proposed in
\cite{schuller0,schuller1}. One advantage of the new formulation
is that it does not rely on the bundle frame language of differential
geometry (DG), making it transparent for non-experts in DG. It will
consist of twice the GR, defined in two separate spaces of the quasi phase
space (its variables are coordinates and velocities, not momenta). Afterward,
both spaces will be connected via the new variational principle.

As will be seen, a consistent formulation can be found and
the results imply the elimination of singular
solutions in GR, in particular there will be no Schwarzschild horizon.
The deviation of the redshift in the new theory from that in standard GR will be determined.
Whether our theory is realized in nature,
has to be verified
by experiment. It is, nevertheless, a valid possibility in the search
of extensions of standard GR.

The paper is structured as follows:
In section II a short review on pseudo-complex variables is given.
In section III the extended formulation of GR is presented and in section IV
the analogue to the standard
Schwarzschild solution will be constructed, which is singularity free.
(We will still refer to it as the Schwarzschild solution.)
In addition, in a subsection, the consequences of the pseudo-complex
Schwarzschild solution are discussed.
In section V the redshift is calculated and
observable deviations from standard GR are estimated.
Finally, in section VI conclusions are
drawn.

Most of the formulation is done in complete analogy to
the book of Adler, Bazin and Schiffer \cite{adler}.
We will simply refer to it at several places. In this way we avoid unnecessary repetition of the literature
and gain, were it is convenient, didactical clarity.

\section{Pseudo-Complex Variables}

Here we give a brief resum\'e on pseudo-complex variables,
helpful to understand the steps presented in this contribution.
The formulas, presented here, can be used without going into the details.
A more profound introduction to pseudo-complex variables is given in
\cite{paper1,paper2}, which can be consulted for better understanding.

The pseudo-complex variables are also known as
{\it hyperbolic} \cite{crumeyrolle1,crumeyrolle2},
{\it hypercomplex} \cite{kantor} or {\it para-complex} \cite{gadea}.
We will continue to use the term pseudo-complex.

The pseudo-complex variables are {\it defined} via

\beqa
X & = & x_1 + I x_2 ~~~,
\eeqa
with $I^2=1$. This is similar to the common complex notation except for
the different behavior of $I$. An alternative presentation is to introduce
the operators

\beqa
\sigma_\pm & = & \frac{1}{2} \left( 1 \pm I \right)  \nonumber \\
\eeqa
with
\beqa
\sigma_\pm^2 & = & 1 ~~~,~~~ \sigma_+\sigma_- = 0 ~~~.
\eeqa
The $\sigma_\pm$ form a so called {\it zero divisor basis},
with the zero divisor defined in mathematical terms by
$\bd{P}^0 = \bd{P}^0_+ \cup \bd{P}^0_-$, with
$\bd{P}^0_\pm=\left\{ X=\lambda \sigma_\pm| \lambda ~\epsilon~ \bd{R} \right\}$.

This basis is used to rewrite the pseudo-complex variables as

\beqa
X & = & X_+ \sigma_+ + X_- \sigma_- ~~~,
\eeqa
with

\beqa
X_\pm = x_1 \pm x_2  ~~~.
\eeqa

The pseudo-complex conjugate of a pseudo-complex variable is

\beqa
X^* & = & x_1 - I x_2 = X_+\sigma_- + X_- \sigma_+ ~~~.
\eeqa
The {\it norm} square of a pseudo-complex variable is given by

\beqa
|X|^2 = XX^ * & = & x_1^2 - x_2^2 ~~~.
\eeqa
This allows for the appearance of a positive, negative and null norm.
Variables with a zero norm are members of the zero-divisor, i.e.,
they are either proportional to $\sigma_+$ or $\sigma_-$.

{\it It is very useful to do all calculations within the zero divisor
basis, $\sigma_\pm$, because all manipulations can be realized
independently in both sectors (because $\sigma_+\sigma_-=0$).}

In each zero divisor component,
differentiation and multiplication can be manipulated in
the same way as with normal variables.
For example, we have \cite{paper2}

\beqa
F(X) & = & F(X_+) \sigma_+ + F(X_-) \sigma_-
\label{f1}
\eeqa
and a product of two functions $F(X)$ and $G(X)$ satisfies

\beqa
F(X)G(X) & = & F(X_+)G(X_+) \sigma_+ + F(X_-)G(X_-) \sigma_-  ~~~.
\label{fg1}
\eeqa

Differentiation is defined as

\beqa
\frac{DF(X)}{DX} & = & \lim_{\Delta X \rightarrow 0}
\frac{F(X+\Delta X)-F(X)}{\Delta X}
~~~,
\eeqa
where $\Delta$ refers from here on to the pseudo-complex 
difference. The $D$ refers to the partial differentiation or
infinitesimal difference.

Finally, we resume some properties of the quasi phase space of a
pseudo-complex four dimensional space-time. It is mainly for completeness
and can be skipped by the non-interested reader.
There are a set of four coordinates $X^\mu$ =
$X_+^\mu \sigma_+ + X_-^\mu \sigma_-$ and four velocities $U^\mu$ =
$U_+^\mu \sigma_+ + U_-^\mu \sigma_-$.
Considering the motion of a mass point,
the coordinates are often
written as $X_\pm^\mu$ = $(x^\mu \pm lu^\mu )$ and the velocities
as $U_\pm^\mu$ = $(u^\mu \pm la^\mu )$, where $x^\mu$, $u^\mu$
= $\frac{dx^\mu}{d\tau}$ ($\tau$ as the eigen-time)
and $a^\mu$ = $\frac{du^\mu}{d\tau}$ are called the
standard coordinates, velocities and
accelerations of the mass point, respectively, along the world line.
There are two separated quasi phase spaces, due to the division in
$\sigma_\pm$ components. One
is built by the pair $(X_+^\mu , U_+^\mu )$ and the other one by
$(X_-^\mu , U_-^\mu )$. All manipulations are done independently
in each subspace of the quasi phase space.
As is well known, canonical transformations are generated by the
members of the algebra of a symplectic group.
Having $n$ coordinates the symplectic group is $Sp(n,R)$.
Therefore, canonical transformations in the extended description of
space-time
exhibit a direct product structure of the type $Sp_+(4,R) \otimes Sp_-(4,R)$,
where $Sp_\pm (4,R)$ is the symplectic group of the four dimensional space.

\section{Formulation of Pseudo-Complex General Relativity}

In a first step, the pseudo-complex metric function is constructed, which
is a pseudo-holomorphic function, i.e., it satisfies
the pseudo-complex Riemann-Cauchy conditions
\cite{paper2}

\beqa
\frac{D g^R_{\mu\nu}}{D X^\lambda_1} & = &
\frac{D g^I_{\mu\nu}}{D X^\lambda_2}  \nonumber \\
\frac{D g^R_{\mu\nu}}{D X^\lambda_2} & = &
\frac{D g^I_{\mu\nu}}{D X^\lambda_1}  ~~~,
\label{cauchy}
\eeqa
where $g^R_{\mu\nu}$ is the pseudo-real and $g^I_{\mu\nu}$ the
pseudo-imaginary component,
with $X_1^\lambda=x^\lambda$ being the pseudo-real part and
$X^\lambda_2$ the pseudo-imaginary part of the 4-coordinate
$X^\lambda = X_1^\lambda+IX_2^\lambda$.

If we would assume that $g_{\mu\nu}$
does only depend on the pseudo-real part $X_1^\lambda = x^\lambda$ of
the coordinate, it would lead us to a non-holomorphic function $g_{\mu\nu}$
(e.g., the first equation in (\ref{cauchy}) would yield zero on the
right hand side, while the left hand side is different from zero).
The condition of $g_{\mu\nu}$ being a pseudo-holomorphic function
is,
therefore, of importance.

Taking into account that differentiating a pseudo-complex function
with respect to a variable follows the same rules as differentiating
a normal function with respect to a variable,
we can try to formulate GR following {\it the same steps} as
indicated in \cite{adler}.
The important difference is that the metric is pseudo-complex.
It is defined as

\beqa
g_{\mu\nu} & = & g^+_{\mu\nu} \sigma_+ + g^-_{\mu\nu} \sigma_- ~~~.
\label{metric}
\eeqa
It is a function of the pseudo-complex space-time variables
$X^\lambda_{\pm}$ and, thus, also a function of the coordinates
and the velocities.

The differential length element squared  is given by

\beqa
d\omega^2 & = & g_{\mu\nu}(X) DX^\mu DX^\nu \nonumber \\
& = & g^+_{\mu\nu}(X_+) DX_+^\mu DX_+^\nu \sigma_+
\nonumber \\
&& + g^-_{\mu\nu}(X_-) DX_-^\mu DX_-^\nu \sigma_-
~~~.
\label{dw2}
\eeqa
The division is within the zero-divisor basis, similar to
Eqs. (\ref{f1}) and (\ref{fg1}).
Interchanging in (\ref{dw2}) the dummy indices $\mu$ and $\nu$ leads
to

\beqa
d\omega^2 & = & g_{\nu\mu}DX^\nu DX^\mu ~=~ g_{\nu\mu}DX^\mu DX^\nu
~~~.
\eeqa
Comparing it to (\ref{dw2}) requires that the metric is symmetric, i.e.,

\beqa
g_{\mu\nu} & = & g_{\nu\mu} ~~~.
\eeqa

{\it Because the $\sigma_\pm$ parts are linearly independent,
it implies that we can formulate a theory of General
Relativity in each of the $\sigma_+$ and the $\sigma_-$
components}. Afterward, we have to connect both sectors, as will be
explained further below. The advantage of the independent formulation
lies in the fact that the GR in each sector will
be analogue to the standard formulation.

For example, a parallel displacement of a pseudo-complex vector $\xi^i$
is given by

\beqa
D\xi^\mu & = & \Gamma_{\nu\lambda}^\mu DX^\nu\xi^\lambda \nonumber \\
& = & \Gamma_{\nu\lambda}^{+~\mu} DX_+^\nu\xi_+^\lambda \sigma_+
+ \Gamma_{\nu\lambda}^{-~\mu} DX_-^\nu\xi_-^\lambda \sigma_- \nonumber \\
& = & d\xi_+^\mu \sigma_+ + d\xi_-^\mu \sigma_-
~~~,
\eeqa
where $DX^\nu$ refers to the change of the pseudo-complex coordinate $X^\nu$
and $\xi^\mu$ are the components of a vector, which is parallel displaced.
The connections $\Gamma_{\nu\lambda}^\mu$ are symmetric in their lower
indices.
The same arguments as in \cite{adler} leads to the now {\it pseudo-complex
Christoffel symbols of the second kind}, starting from the
condition that the pseudo-complex
line-squared element $d\omega^2$ is required
to be invariant under the transformation of the coordinates.
The pseudo-complex {\it Christoffel symbols
of the second kind} are given by

\beqa
\Gamma_{\mu\nu}^\lambda & = &
-\left\{
\begin{array}{ccc}
& \lambda & \\
\nu && \mu
\end{array}
\right\}
\nonumber \\
& = &
-\left\{
\begin{array}{ccc}
& \lambda & \\
\nu && \mu
\end{array}
\right\}_+ \sigma_+
-\left\{
\begin{array}{ccc}
& \lambda & \\
\nu && \mu
\end{array}
\right\}_- \sigma_-
~~~,
\eeqa
which can be written in terms of {\it Christophel symbols of
the first kind} \cite{adler} as

\beqa
\Gamma_{\mu\nu}^{\pm~\lambda}
& = &
-\left\{
\begin{array}{ccc}
& \lambda & \\
\nu && \mu
\end{array}
\right\}_{\pm}
~=~ -g^{\lambda\kappa}\left[ \nu\mu,\kappa \right]_\pm
~~~.
\label{first}
\eeqa
The {\it Christoffel symbol of the first kind} are defined as \cite{adler}

\beqa
\left[ \mu\nu,\kappa \right] & = &
\frac{1}{2} \left(
\frac{D g_{\mu\kappa}}{D X^\nu}
+ \frac{D g_{\nu\kappa}}{D X^\mu} - \frac{D g_{\mu\nu}}{D X^\kappa} \right)
~~~.
\label{first2}
\eeqa
The expression $\frac{D g_{\mu\lambda}}{D X^\nu}=g_{\mu\lambda|\nu}$ denotes the pseudo-complex
derivative of $g_{\mu\lambda}$ with respect to $X^\nu$.

The 4-derivative of a contravariant vector is given by

\beqa
\xi^\mu_{||\nu} & = & \xi^\mu_{|\nu} +
\left\{
\begin{array}{ccc}
& \mu & \\
\nu && \lambda
\end{array}
\right\}
\xi^\lambda
\nonumber \\
& = & \left(
\xi^\mu_{+|\nu} +
\left\{
\begin{array}{ccc}
& \mu & \\
\nu && \lambda
\end{array}
\right\}_+
\xi_+^\lambda \right) \sigma_+
\nonumber \\
&& + \left(
\xi^\mu_{-|\nu} +
\left\{
\begin{array}{ccc}
& \mu & \\
\nu && \lambda
\end{array}
\right\}_-
\xi_-^\lambda \right) \sigma_-
~~~,
\eeqa
where $\xi^\mu_{|\nu} = \frac{D\xi^\mu}{DX^\nu}$.
The rules for deriving covariant vectors and tensors can be
directly copied from \cite{adler}.

{\it An important point is that in this new formulation the
4-divergence of the metric
will again be zero!} To show this, we copy the arguments,
as given in \cite{adler}, chapter 3. We have

\beqa
g^\pm_{\mu\nu |\lambda}-g^\pm_{\mu\kappa}
\left\{
\begin{array}{ccc}
& \kappa & \\
\nu && \lambda
\end{array}
\right\}_\pm
& = & [\mu\lambda,\nu]_\pm ~~~,
\label{man1}
\eeqa
where the symmetry property of the metric tensor was used.
Eq. (\ref{man1}) is proved by substituting the Christoffel symbol
of the second kind (\ref{first}) and using the definition of the
Christoffel symbol of the first kind (\ref{first2}).

Using Eq. (\ref{man1}), the divergence of $g^\pm_{\mu\nu}$ can be
rewritten as

\beqa
g^\pm_{\mu\nu ||\lambda} & = & g^\pm_{\mu\nu |\lambda} -
\left\{
\begin{array}{ccc}
& \kappa & \\
\nu && \lambda
\end{array}
\right\}_\pm
g^\pm_{\mu\kappa} -
\left\{
\begin{array}{ccc}
& \kappa & \\
\mu && \lambda
\end{array}
\right\}_\pm
g^\pm_{\kappa\nu} \nonumber \\
& = & [\mu\lambda,\nu]_\pm - g^\pm_{\kappa\nu}
\left\{
\begin{array}{ccc}
& \kappa & \\
\mu && \lambda ~~~.
\end{array}
\right\}_\pm
\eeqa
Utilizing the definition of the Christoffel symbol of the second kind
(see above), this
expression is identical to zero. Thus, also the 4-divergence of the
pseudo-complex metric is zero:

\beqa
g_{\mu\nu ||\lambda} & = & g^+_{\mu\nu ||\lambda} \sigma_+
+ g^-_{\mu\nu ||\lambda} \sigma_- ~=~ 0 ~~~,
\eeqa
or equivalently

\beqa
g^\pm_{\mu\nu ||\lambda} & = & 0 ~~~,
\eeqa
where the derivative is now with respect to the coordinates
$X_\pm^\lambda$.

{\it This result is very important}: It is a necessary requirement for the principle
of General Relativity. (The tensor $g_{\mu\nu}$ is invariant
under the action of the four-derivative, thus it is invariant
under a parallel displacement. The same holds trivially for
the operator $I$, because it is constant. Thus, we have an almost
product structure \cite{jano}.)

This also leads to two different kinds of four derivatives, one for the
$\sigma_+$ and the other one for the $\sigma_-$ component.

In order to proceed further, we need to introduce an important difference
to the treatment of standard GR. It is the {\it change
in the variational principle}:
Up to now, it seems that we have only a double, parallel, formulation of
GR, one in the $\sigma_+$ and the other one in the $\sigma_-$ component.
In the next step we show how both zero-divisor components
are {\it linked together}. We will follow a suggestion given in
\cite{paper1,paper2,schuller0,schuller1}: \\
Following the Lagrange formulation and
denoting by $L$ the Lagrangian within an integral, we have the action

\beqa
S & = & \int L d\tau ~~~.
\eeqa
The {\it variational procedure is now modified to} \cite{paper2}

\beqa
\delta S & = & \delta \int L d\tau ~\epsilon~ {\bd P}^0 ~~~,
\eeqa
with ${\bd P}^0$ being the zero divisor (numbers linear in $\sigma_+$ or
$\sigma_-$ only \cite{paper2}). One argument is that the zero divisor
branch consists of numbers which have a zero norm and in this sense it
represents a generalized zero.

To illustrate it more, suppose we would require that the variation
of the action is exactly zero, then one gets that
$\delta S$ = $\delta S_+ \sigma_+$ +
$\delta S_- \sigma_-$ = 0, or $\delta S_\pm = 0$.
In other words, one would obtain simply a double formulation
of GR. However, if it is required that the
variation of the action is within the zero-divisor branch, then both
components are linked and only then it makes sense to obtain a
new, modified theory of General Relativity.

The variation of the action leads to

\beqa
\frac{D}{Ds}\left( \frac{DL}{DX^\mu} \right) - \frac{DL}{DX^\mu} ~\epsilon~
{\bd P}^0 ~~~,
\eeqa
with $s$ as some curve parameter, which can be the eigen-time $\tau$.
Note, that the right hand side has to be in the zero divisor, i.e.,
it is proportional either to $\xi_\mu\sigma_-$ or $\xi_\mu\sigma_+$,
with $\xi_\mu$ a real or normal complex number or function. These
$\xi$'s can be used as an additional freedom to fix solutions of
the equations of motion and {\it will play a crucial role}.

As a Lagrangian one can use the length element, which leads to the equation
of geodesics (in fact two, for each component in the zero-divisor basis):

\beqa
{\ddot X}^\mu +
\left\{
\begin{array}{ccc}
& \mu & \\
\nu && \lambda
\end{array}
\right\}
{\dot X}^\nu {\dot X}^\lambda & \epsilon & {\cal P}^0 ~~~.
\label{eq-m-x}
\eeqa

This, however, assumes a test-particle description, as explained in
Ref. \cite{adler}. Expressing $L$ in terms of a curvature tensor,
which is independent to the use of a test particle, we obtain
\cite{adler} for a matter free space

\beqa
G_{\mu\nu} & = & {\cal R}_{\mu\nu} - \frac{1}{2}g_{\mu\nu}{\cal R} ~\epsilon~
{\cal P}^0 ~~~,
\label{einsteintensor}
\eeqa
modified by the zero divisor on the right hand side.
$G_{\mu\nu}$ is the pseudo-complex Einstein tensor, ${\cal R}_{\mu\nu}$ is the
Ricci tensor, defined in the same way as in standard GR, with the difference
that now it is pseudo-complex. The ${\cal R}$ is the Riemann curvature
(also known as {\it scalar curvature}).
The Lagrangian used has the form $L$ =
$\sqrt{-g}{\cal R}$, with $g$ being the determinant of the metric tensor.
In standard GR the equations of motion reduce, for a matter free space,
to the Ricci tensor equal to
zero, i.e. ${\cal R}_{\mu\nu}=0$. In our procedure, this is extended to

\beqa
{\cal R}_{\mu\nu} & = & {\cal R}^+_{\mu\nu} \sigma_+
+ {\cal R}^-_{\mu\nu} \sigma_-
~ \epsilon ~ {\cal P}^0 ~~~.
\label{eq-motion}
\eeqa
Comparing it with (\ref{einsteintensor}) leads to

\beqa
{\cal R} & = & {\cal R}_+ \sigma_+ + {\cal R}_- \sigma_- ~=~ 0 ~~~.
\label{r=0}
\eeqa
{\it This is an important result. It means that the
space has still a local zero scalar curvature}. This gives us an additional
and necessary
relation which will fix the functions appearing on the right hand side of the
equation of motion.

Some words have to be said about the integrability of the system: The
assumption that ${\cal R}_{\mu\nu}$ is not exactly zero but in the
zero divisor basis (e.g., proportional to $\sigma_-$) implies that a
pseudo-complex vector parallel displaced along two curves leads to,
in general, two different vectors. However, the two vectors differ only by
a component within the zero divisor basis. If
$\xi_1^\mu$ denotes the vector obtained after parallel displacement along
a curve $C_1$ and $\xi_2^\mu$ is the vector obtained after parallel
displacement along the curve $C_2$, then $\xi_1^\mu - \xi_2^\mu$
$\epsilon$ ${\cal P}^0$. The difference vector has zero norm.
{\it In the pseudo-complex space we, therefore, consider all
vectors to be equivalent which differ only by a component in the zero
divisor basis, where by convention we will choose it to be proportional to
$\sigma_-$}.
This proposal is a generalized definition of integrability. As we will
see in section IV, it leads finally to a standard description of GR, were
integrability is assured, while the effect of the pseudo-complex
description is surviving through a particular contribution in the metric
components. Thus, the description presented here, shows a possible
alternative description of GR by making a detour to pseudo-complex numbers,
an idea not thought before.

{\it As just defined, in the equation of motion we will
assume for convenience}
\cite{paper1,paper2}, {\it that the right hand side is
proportional to $\sigma_-$}. The assumption of being proportional
to $\sigma_+$ leads to a symmetric description.

\subsection{Further properties of the metric}

In this section we discuss further properties related to the metric,
useful for subsequent calculations.

In the zero-divisor basis, the metric is given by Eq. (\ref{metric}).
The product with its inverse, $g^{\mu\nu}$, yields

\beqa
g^{\mu\nu} g_{\nu\lambda} & = & g_+^{\mu\nu} g^+_{\nu\lambda} \sigma_+
+ g_-^{\mu\nu} g^-_{\nu\lambda} \sigma_- \nonumber \\
& = & \delta_{\mu\lambda} \left( \sigma_+ + \sigma_- \right)
~=~ \delta_{\mu\lambda} ~~~.
\eeqa
This choice contains two, in general, different metrics in the $\sigma_\pm$
parts.

The pseudo-real and pseudo-imaginary part of the metric is related
to $g^+_{\mu\nu}$ and $g^-_{\mu\nu}$ by

\beqa
g^R_{\mu\nu} & = & \frac{1}{2} \left( g^+_{\mu\nu} + g^-_{\mu\nu} \right)
~=~ g^0_{\mu\nu}
\nonumber \\
g^I_{\mu\nu} & = & \frac{1}{2} \left( g^+_{\mu\nu} - g^-_{\mu\nu} \right)
~=~ h_{\mu\nu}
~~~,
\eeqa
where we introduce a new, more convenient notation in terms of an
{\it average} metric $g^0_{\mu\nu}$ and a {\it difference} metric
$h_{\mu\nu}$. This leads to

\beqa
g^\pm_{\mu\nu} & = & g^0_{\mu\nu} \pm  h_{\mu\nu} ~~~.
\label{l3}
\eeqa

The metric $g^\pm_{\mu\nu}$ lowers the index of $X_\pm^\mu$
and $P_\pm^\mu$, while
$g_\pm^{\mu\nu}$ raises the ones of $X^\pm_\mu$ and $P^\pm_\mu$, i.e,

\beqa
X^\pm_\mu & = & g^\pm_{\mu\nu} X_\pm^\nu \nonumber \\
X_\pm^\mu & = & g_\pm^{\mu\nu} X^\pm_\nu
\label{relat}
\eeqa
or equivalently

\beqa
x_\mu \pm l u_\mu & = & g^\pm_{\mu\nu} \left( x^\nu \pm lu^\nu \right)
\nonumber \\
x^\mu \pm l u^\mu & = & g_\pm^{\mu\nu} \left( x_\nu \pm lu_\nu \right)
\label{relat2}
\eeqa
and similar for $P_\pm^\mu$ =$p_\mu \pm l f_\mu$ and
$P^\pm_\mu$ = $p^\mu \pm l f^\mu$, with $p^\mu$ as the linear
momentum and $f^\mu$ being an object with the units of a force. Note,
that one has to apply $g^\pm_{\mu\nu}$
on $X_\pm^\mu$ and not separately on $x^\mu$ and $u^\mu$.

From the former equations (after subtracting and adding the first
two equations in (\ref{relat2}), in order
to solve for $x_\mu$ and $p_\mu$) we obtain

\beqa
x_\mu & = & \frac{1}{2} \left( g^+_{\mu\nu} + g^-_{\mu\nu} \right) x^\nu
+ l\frac{1}{2} \left( g^+_{\mu\nu} - g^-_{\mu\nu} \right) u^\nu
\nonumber \\
& = & g^0_{\mu\nu} x^\nu + lh_{\mu\nu} u^\nu \nonumber \\
lu_\mu & = & \frac{1}{2} \left( g^+_{\mu\nu} - g^-_{\mu\nu} \right) x^\nu
+ l\frac{1}{2} \left( g^+_{\mu\nu} + g^-_{\mu\nu} \right) u^\nu
\nonumber \\
& = & lg^0_{\mu\nu} u^\nu + h_{\mu\nu} x^\nu
~~~.
\eeqa
One feature is that the raising and lowering of the indices
can be applied only via a metric in the zero divisor components
of $X^\mu$ ($X_\mu$), i.e., the {\it individual expressions
of the coordinate} ($x^\mu$, $x_\mu$) {\it and velocities} ($u^\mu$, $u_\mu$)
{\it are not contra- and covariant vectors any more}. The consequences
of this have still to be explored. The exception happens in the limit
when $l$ and $h_{\mu\nu}$
are zero, then $g^0_{\mu\nu}$ ($g_0^{\mu\nu}$) lower (raise) the
components of the space-time and 4-velocity components.

The invariant generalized length element is given by
Eq. (\ref{dw2}).
{\it Because it is an observables, it is required to be
pseudo-real.}. Thus, 

\beqa
d\omega^{*~2} & = & d\omega^2 ~~~.
\eeqa
From this condition we obtain the following relation

\beqa
& g^+_{\mu\nu}(X_+)DX_+^\mu DX_+^\nu \sigma_+ +
g^-_{\mu\nu}(X_-)DX_-^\mu DX_-^\nu \sigma_ - & \nonumber \\
& = & \nonumber \\
& g^+_{\mu\nu}(X_+)DX_+^\mu DX_+^\nu\sigma_- +
g^-_{\mu\nu}(X_-)DX_-^\mu DX_-^\nu \sigma_ + &
~~~,
\eeqa
or

\beqa
g^+_{\mu\nu}(X_+)DX_+^\mu DX_+^\nu & = & g^-_{\mu\nu}(X_-)DX_-^\mu DX_-^\nu
~~~.
\eeqa

Expressing the $X_\pm^\mu$ in terms of $x^\mu$ and $u^\mu$ leads to

\beqa
h_{\mu\nu} \left(dx^\mu dx^\nu + l^2 du^\mu du^\nu \right) && \nonumber \\
+lg^0_{\mu\nu} \left(dx^\mu du^\nu + du^\mu dx^\nu \right) & = & 0 ~~~.
\label{hmunu}
\eeqa
This is a generalized version for the "orthogonality" of $dx$ and $du$.
For the special case of a flat space ($h_{\mu\nu}=0$ and $g^0_{\mu\nu}=
\eta_{\mu\nu}$) we arrive at the standard relation of $dx_\mu du^\mu =0$.
(We use the signature $\eta_{\mu\nu}=(+,-,-,-)$.)

With Eq. (\ref{hmunu}), the $d\omega^2$ acquires the form

\beqa
d\omega^2 & = & g^0_{\mu\nu} \left(dx^\mu dx^\nu +
l^2 du^\mu du^\nu \right) \nonumber \\
&& +lh_{\mu\nu} \left(dx^\mu du^\nu + du^\mu dx^\nu \right)
~~~.
\label{gmunu}
\eeqa

The new invariant length element can be simplified if only terms
up to the order in $l^0$ are considered.
It leads to

\beqa
d\omega^  2 & \approx & g^0_{\mu\nu} dx^\mu dx^\nu
~~~.
\label{domega}
\eeqa
As one can see, applying the above approximations {\it reduces the
generalized length element} $d\omega^2$ {\it to the standard
form known for} $ds^2$. This can be understood, considering that
$d\omega^2$ is related to $ds^2$ by a factor of the type
$\left( 1-l^2 a^2 \right)$, with $l$ as the minimal length parameter
and $a$ for the acceleration
\cite{caneloni,brandt1,brandt2,brandt3,beil1,beil2,beil3,beil4}.
Taking into account only terms proportional to $l^0$ reduces this factor
to 1.
The important point is that $g^0_{\mu\nu}$ now depends on the
differences between $g^\pm_{\mu\nu} (X_\pm )$, which contains remnant
contributions from the pseudo-complex description. Note, that within this
approximation we will return to the standard description of GR. What we will
not discuss here are the contributions generated by corrections due to the
minimal length element. We refer to a later publication.

This expression of the line element will play a crucial role in the next
section, devoted to the discussion of the Schwarzschild solution within the
pseudo-symplectic formulation.

\section{Pseudo-complex Schwarzschild solution}

Again, we follow closely the book \cite{adler}, chapter 6.

In a first step, we deduce the pseudo-complex Christoffel symbols
of the second kind, using
the method described in \cite{adler}.

The length element is deduced by requiring that it is invariant under
$DX^0 \rightarrow -DX^0$, $D\theta \rightarrow -D\theta$ and
$D\phi \rightarrow -D\phi$. Following the same steps as in chapter 6 of
\cite{adler}, the resulting pseudo-complex length element is given by

\beqa
d\omega^2 & = & A\left( DX_0\right)^2 -B \left(DR\right)^2 -
R^2 \left( \left(D\theta\right)^2 + sin^2 \theta \left(D\phi\right)^2
\right)
~~~.
\eeqa
The $A_\pm$ and $B_\pm$ functions are positive definite and one can write
them as $A_\pm=e^{\nu_\pm (R_\pm )}$ and $B_\pm=e^{\lambda_\pm (R_\pm )}$.
Because of this, we can also write $A=A_+\sigma_+ + A_- \sigma_-$ and
$B=B_+\sigma_+ + B_- \sigma_-$ as $A=e^{\nu (R )}$ and $B=e^{\lambda (R )}$
respectively.
The same result is obtained when one parts from
the line element in each zero divisor component ($d\omega_\pm^2$)
and imposes the same symmetry conditions to the corresponding variables.
It is nothing but proceeding in each zero divisor component as in standard
GR.

In order to deduce the Christoffel symbols, we recur to a trick by
first determining the equation of geodesics. We use as a
variational principle

\beqa
\delta \int \left[ A{\dot X}_0^2 -B{\dot R}^2 -R^2 \left( {\dot \theta}^2 +
sin^2 \theta {\dot \phi}^2 \right) \right]dp & \epsilon & {\cal P}^0 ~~~,
\label{vari}
\eeqa
where a dot above a variables indicates its derivative with respect to $p$,
the curve parameter along the world line of  particle (for example, $p$
can be the eigen-time $\tau$, the arc length $s$, etc.).
The variation leads to equations of motion of the form of (\ref{eq-m-x}).
By our convention we will require that the $\sigma_-$ component
is different from zero. The right hand side
of (\ref{eq-m-x}) can be a function in the
pseudo-complex variables.

These equations have to be compared with those obtained from
(\ref{vari}), which are

\beqa
{\ddot X}^0 + \nu^\prime {\dot R} {\dot X}^0 & = & \xi^0\sigma_- \nonumber \\
{\ddot R} + \frac{1}{2}\lambda^\prime {\dot R}^2 +
\frac{1}{2}\nu^\prime e^{\nu-\lambda}
({\dot X}^0)^2 - e^{-\lambda} R {\dot \theta}^2  && \nonumber \\
- Rsin^2\theta {\dot \phi}^2
e^{-\lambda} & = & \xi_R \sigma_- \nonumber \\
{\ddot \theta} + \frac{2}{R} {\dot \theta} {\dot R} -sin\theta cos\theta
{\dot \phi}^2 & = & \xi_\theta \sigma_- \nonumber \\
{\ddot \phi} + 2cot\theta {\dot \phi}{\dot \theta} + \frac{2}{R}
{\dot R} {\dot \phi} & = & \xi_\phi \sigma_- ~~~,
\label{eq-R}
\eeqa
for $X^0$, $R$, $\theta$ and $\phi$ respectively.
A prime indicates the derivative with respect to $R$, e.g.,
$\nu^\prime = \frac{D\nu}{DR}$.
On the right hand side of each equation we put an element of the
zero-divisor, such that it is proportional to $\sigma_-$, following our
convention. The $\sigma_\pm$ components of the $\lambda$ and $\nu$
are functions in the variables $X_\pm^\mu$.
Comparing (\ref{eq-R}) with (\ref{eq-m-x}) gives us the
Christoffel symbols of the second kind.
The non-zero Christoffel symbols are given by

\beqa
\left\{
\begin{array}{ccc}
& 0 & \\
1 && 0
\end{array}
\right\} &=& \frac{1}{2}\nu^\prime
~=~
\left\{
\begin{array}{ccc}
& 0 & \\
0 && 1
\end{array}
\right\}
\nonumber \\
\left\{
\begin{array}{ccc}
& 1 & \\
0 && 0
\end{array}
\right\} &=& \frac{1}{2}\nu^\prime e^{\nu-\lambda} \nonumber \\
\left\{
\begin{array}{ccc}
& 1 & \\
1 && 1
\end{array}
\right\} &=& \frac{1}{2}\lambda^\prime
\nonumber \\
\left\{
\begin{array}{ccc}
& 1 & \\
2 && 2
\end{array}
\right\} &=& -R e^{-\lambda} \nonumber \\
\left\{
\begin{array}{ccc}
& 1 & \\
3 && 3
\end{array}
\right\} &=& -R sin^2 \theta e^{-\lambda}
\nonumber \\
\left\{
\begin{array}{ccc}
& 2 & \\
2 && 1
\end{array}
\right\} &=& \frac{1}{R} ~= ~
\left\{
\begin{array}{ccc}
& 2 & \\
1 && 2
\end{array}
\right\}  \nonumber \\
\left\{
\begin{array}{ccc}
& 2 & \\
3 && 3
\end{array}
\right\} &=& - sin \theta ~ cos \theta
\nonumber \\
\left\{
\begin{array}{ccc}
& 3 & \\
2 && 3
\end{array}
\right\} &=& cot \theta ~=~
\left\{
\begin{array}{ccc}
& 3 & \\
3 && 2
\end{array}
\right\} \nonumber \\
\left\{
\begin{array}{ccc}
& 3 & \\
1 && 3
\end{array}
\right\} &=& \frac{1}{R} ~=~
\left\{
\begin{array}{ccc}
& 3 & \\
3 && 1
\end{array}
\right\}  ~~~.
\label{chris}
\eeqa

In the next step we use the proposed equation of motion, as given in
Eq. (\ref{eq-motion}) above, with the subsidiary condition
for the curvature
(${\cal R}=0$, see Eq. (\ref{r=0})). For that, we remind
on the structure of the metric, which is

\beqa
\left(
\begin{array}{cccc}
e^{\nu (R)} & 0 & 0 & 0 \\
0 & -e^{\lambda (R)} & 0 & 0 \\
0 & 0 & -R^2 & 0 \\
0 & 0 & 0 & -R^2 sin^2 \theta
\end{array}
\right)                       
\label{metric3}
\eeqa
and its determinant is

\beqa
g & = & -e^{\nu + \lambda} R^4 sin^2 \theta
~~~.
\eeqa
For its logarithm we get

\beqa
ln\sqrt{-g} & = & \frac{\nu + \lambda}{2} + 2 ln R + ln|sin\theta| ~~~.
\label{log-g}
\eeqa
The Ricci tensor is of the form

\beqa
{\cal R}_{\mu\nu} & = &
\left\{
\begin{array}{ccc}
& \beta & \\
\beta && \nu
\end{array}
\right\}_{|\mu} -
\left\{
\begin{array}{ccc}
& \beta & \\
\mu && \nu
\end{array}
\right\}_{|\beta} \nonumber \\
&& +
\left\{
\begin{array}{ccc}
& \beta & \\
\tau && \mu
\end{array}
\right\}
\left\{
\begin{array}{ccc}
& \tau & \\
\beta && \nu
\end{array}
\right\}   \nonumber \\
&& -
\left\{
\begin{array}{ccc}
& \beta & \\
\tau && \beta
\end{array}
\right\}
\left\{
\begin{array}{ccc}
& \tau & \\
\mu && \nu
\end{array}
\right\}  ~~~.
\label{ricci}
\eeqa

Using (\ref{chris}) and
the explicit expressions for the Christoffel symbols of Eq.
(\ref{chris}), we obtain for the
${\cal R}_{\mu\mu}$ components

\beqa
{\cal R}_{00} & = &
-\frac{e^{\nu-\lambda}}{2} \left( \nu^{\prime\prime}
+\frac{\nu^{\prime 2}}{2} - \frac{\lambda^\prime \nu^\prime}{2}
+\frac{2\nu^\prime}{R} \right) \nonumber \\
{\cal R}_{11} & = &
\frac{1}{2} \left( \nu^{\prime\prime}
+\frac{\nu^{\prime 2}}{2} - \frac{\lambda^\prime \nu^\prime}{2}
-\frac{2\lambda^\prime}{R} \right) \nonumber \\
{\cal R}_{22} & = &
\left( e^{-\lambda} R \right)^\prime - 1 \nonumber \\
{\cal R}_{33} & = &
sin^2\theta ~ \left[ \left( e^{-\lambda} R \right)^\prime - 1 \right] ~~~.
\label{rmn}
\eeqa

With that, the equations for ${\cal R}_{00}$ and ${\cal R}_{11}$
are resulting:

\beqa
\nu^{\prime\prime} + \frac{1}{2} \nu^{\prime  2} - \frac{1}{2} \lambda^\prime \nu^\prime
+ \frac{2\nu^\prime}{R} & = & \xi_0 \sigma_-
\nonumber \\
\nu^{\prime\prime} + \frac{1}{2} \nu^{\prime 2} - \frac{1}{2} \lambda^\prime \nu^\prime
- \frac{2\lambda^\prime}{R} & = & \xi_1 \sigma_-  ~~~.
\label{r00r11}
\eeqa
On the left hand side only functions in $R$ appear. Hence, the
$\xi_0$ and $\xi_1$ functions depend on $R$ only, too.
(In fact, because $f(R)\sigma_-$ = $f(R_-)\sigma_-$, these functions
depend only on $R_-$).

Subtracting both equations and utilizing $R\sigma_-=R_-\sigma_-$
yields:

\beqa
\nu^\prime + \lambda^\prime & = &
\frac{1}{2} R_- \left( \xi_0(R_-)-\xi_1(R_-) \right) \sigma_-
~~~.
\label{nula}
\eeqa
The length element is a function in $\lambda$ and $\nu$, i.e.,

\beqa
d\omega^2 & = & e^\nu (DX^0)^2 -e^\lambda \left( DR\right)^2
\nonumber \\
&& -R^2 \left(
\left( D\theta\right)^2 + sin ^2\theta \left( D\phi\right)^2 \right) ~~~.
\eeqa
The solution (\ref{nula}) implies that

\beqa
e^{-\lambda} & = & e^{\nu-\int\frac{R(\xi_0 -
\xi_1)}{2}dR_-}\sigma_- \nonumber \\
& = & e^{\nu_+} \sigma_+ + e^{\nu_- -
\int\frac{R_-(\xi_0 - \xi_1)}{2}dR_-} \sigma_- ~~~.
\label{nula2}
\eeqa

{\it This gives us a restriction on $\xi_0$ and $\xi_1$}: Suppose,
we get a sensible result for $\lambda$, which goes to zero for
large values of $r$, such that $e^\lambda \rightarrow 1$
and $e^\nu \rightarrow 1$.
For that, we note that $R_\pm = r \pm l{\dot r}$
(${\dot r}=\frac{dr}{d\tau}$ is the radial velocity), i.e., its pseudo-real component is $r$,
while its pseudo-imaginary component is $l{\dot r}$, and $R_+$ ($R_-$)
are the sum (difference) of these components.
Then Eq. (\ref{nula2}) implies that $e^{\nu_-}$
tends for large $r$ to a value different from one.
The limit 1 is only achieved if the integrand is set to zero, i.e., 

\beqa
\xi_0 & = & \xi_1 ~~~,
\label{xi0}
\eeqa
i.e., $\nu^\prime = -\lambda^\prime$.

Substituting this into the second equation of (\ref{r00r11}),
which originates from ${\cal R}_{11}$, we obtain

\beqa
\lambda^{\prime\prime} - \lambda^{\prime 2} + \frac{2\lambda^\prime}{R} & = &
-\xi_1 \sigma_-  \nonumber \\
& = & -\frac{e^{\lambda}}{R} \left( R e^{-\lambda} \right)^{\prime\prime}
~~~.
\label{r11}
\eeqa
The last expression leads to

\beqa
\left(Re^{-\lambda}\right)^\prime & = & const +\int R_-
e^{-\lambda_-} \xi_1(R_-) dR_- \sigma_- ~~~.
\label{eqa}
\eeqa

Using the expression for ${\cal R}_{22}$ yields

\beqa
\left( e^{-\lambda} R \right)^\prime & = &
\left( e^{-\lambda_+} R_+ \right)^\prime \sigma_+ +
\left( e^{-\lambda_-} R_- \right)^\prime \sigma_-    \nonumber \\
& = & 1 + \xi_2 \sigma_-
~~~,
\label{eqb}
\eeqa
which determines the right hand side of (\ref{eqa}).
The main conclusion is that the expression on the left
hand side of in Eq. (\ref{eqb}) is proportional to 1 plus
a function in $R_-$ (i.e., $\xi_2 (R_-)$
times $\sigma_-$, being an element of the zero divisor branch).

Integrating Eq. (\ref{eqb}) yields

\beqa
e^{-\lambda}R & = & R - 2{\cal M} +
\int \xi_2(R_-) dR_-\sigma_- ~~~,
\eeqa
where $-2{\cal M}= -2(M_+\sigma_+ + M_- \sigma_-)$
is a pseudo-complex integration constant.

For the $\sigma_+$ part ($\lambda \rightarrow \lambda_+$ and $R
\rightarrow R_+$), the equation is equivalent to the one in the
book of Adler et al., i.e.,

\beqa
e^{-\lambda_+} & = & 1- \frac{2M_+}{R_+} ~~~.
\eeqa
Let us, therefore, restrict to the
$\sigma_-$ part only. The $\sigma_-$ part reads

\beqa
e^{-\lambda_-} & = & 1 - \frac{2M_-}{R_-} +
\frac{1}{R_-}\int \xi_2(R_-) dR_- ~~~.
\label{lala}
\eeqa

We have to substitute this expression into
equation (\ref{r11}), which resulted from ${\cal R}_{11}$, i.e.,

\beqa
\lambda_-^{\prime\prime} - \lambda_-^{\prime 2} +
\frac{2\lambda_-^\prime}{R_-} & = & -\frac{e^{\lambda_-}}{R_-}
\left( R_- e^{-\lambda_-} \right)^{\prime\prime}
\nonumber \\
& = & -\xi_1
~~~.
\label{54}
\eeqa
This leads to

\beqa
\left( R_- e^{-\lambda_-} \right)^{''} & = & \xi_1(R_-)R_-
e^{-\lambda_-(R_-)} ~~~.
\eeqa
Utilizing Eq. (\ref{eqb}), we arrive at

\beqa
\left(1+\xi_2(R_-)\right)^\prime & = & \xi_2^\prime(R_-) ~=~
R_-\xi_1(R_-) e^{-\lambda_-(R_-)} ~~~.
\eeqa
Using Eq. (\ref{lala})
on the right hand side, we get

\beqa
\xi_2^\prime & = & \xi_1\left[ R_- - 2M_- +\int \xi_2
dR_- \right] ~~~.
\label{xi1-xi2}
\eeqa

Note, that the dimension of $\xi_0=\xi_1$ is one over length
squared (see Eq. (\ref{54})). In contrast $\xi_2$ has
no dimension (see Eq. (\ref{eqb})).

It remains to see what the other relations
${\cal R}_{\mu\nu}~\epsilon~{\cal P}^0$,
for $(\mu ,\nu )=(3,3)$ and also for $\mu\ne\nu$ give, considering only the
$\sigma_-$ part.

Following again the book of Adler et al., for $(\mu ,\nu )=(3,3)$,
using Eq. (\ref{rmn}), we get

\beqa
sin^2\theta \left[\left( R_- e^{-\lambda_-}\right)^\prime -1
\right] & = & \xi_3 ~~~,
\label{r33}
\eeqa
where according to Eq. (\ref{eqb})
the parenthesis $\left[...\right]$ on the left side is just $\xi_2$.
Therefore, $\xi_3$ is of the form

\beqa
\xi_3 & = & \xi_2 sin^2\theta ~~~.
\label{xi3}
\eeqa
{\it Thus we arrive at a set of relations for the $\xi_k$ ($k=0,1,2,3$).
All can be expressed in terms of, e.g., $\xi_1$.}

As in Ref. \cite{adler}, all other components of $R_{\mu\nu}$, with
$\mu\ne\nu$ are identically zero, which is proved using the explicit
expressions (\ref{ricci}) of the Ricci tensor and the list of the non-zero
components (\ref{chris}) of the Christoffel symbols of the second kind..

The $\sigma_-$ component of $e^{-\lambda}$ now reads (see (\ref{54}))

\beqa
e^{-\lambda_-} & = & 1-\frac{2M_-}{R_-} + \frac{1}{R_-}\int \xi_2 dR_-
\nonumber \\
& = & 1-\frac{2M_-}{R_-} + \frac{\Omega}{R_-}
~~~,
\label{int-a}
\eeqa
where the $\Omega (R_-)$ function is defined as

\beqa
\Omega & = & \int \xi_2 dR_- ~~~.
\label{omeg}
\eeqa

{\it The next and last step consists in applying the condition ${\cal R}=0$,
which relates
the $\xi_1$ with the $\xi_2$ function}. Using the metric
(\ref{metric3}) and that
${\cal R}=g^{\mu\nu}{\cal R}_{\mu\nu}$, being a scalar in the
$\sigma_\pm$ components, we obtain

\beqa
{\cal R} & = & e^{-\nu}{\cal R}_{00}-e^{-\lambda} {\cal R}_{11} -
\frac{1}{R^2}{\cal R}_{22}
-\frac{1}{R^2sin^2\theta} {\cal R}_{33} \nonumber \\
& = & \left( -e^{-\lambda} \xi_1 - \frac{2}{R^2} \xi_2 \right) \sigma_-
~=~ \left( -e^{-\lambda_-} \xi_1 - \frac{2}{R_-^2} \xi_2 \right) \sigma_-
\nonumber \\
& = & 0
~~~,
\label{rel-xi}
\eeqa
where we have used on one side the relations between the $\xi$-functions and
the components ${\cal R}_{\mu\nu}$ of the Ricci tensor
(\ref{rmn}), (\ref{r00r11}), (\ref{eqb}) , (\ref{r33}) and on the other side the relations
between the $\xi_\mu$ functions (\ref{xi0}), (\ref{xi3}).
The dependence of the ${\cal R}_{\mu\mu}$ components
on the $\xi$ functions, which can be deduced from the equations
in (\ref{rmn}), are

\beqa
{\cal R}_{00} & = & -\frac{1}{2}e^{\nu-\lambda}\xi_0 \sigma_- ~~,~~
{\cal R}_{11} ~= ~ \frac{1}{2}\xi_1 \sigma_- \nonumber \\
{\cal R}_{22} & = & \xi_2 \sigma_- ~~,~~
{\cal R}_{33} ~=~ \xi_3 \sigma_- ~~~.
\eeqa
We obtain from (\ref{rel-xi})

\beqa
\xi_1 & = & -\frac{2e^{\lambda_-}}{R_-^2}\xi_2 ~~~.
\eeqa
The result is substituted into  (\ref{xi1-xi2}), yielding

\beqa
\xi_2^\prime & = & -\frac{2}{R_-}\xi_2 ~~~.
\eeqa
which is a differential equation for $\xi_2$, with the solution

\beqa
\xi_2 & = & \frac{-B}{R_-^2} ~~~,
\label{xi2}
\eeqa
where $B$ is an integration constant. The minus sign
is for convenience and can be understood
further below. For the function $\Omega$ (see (\ref{omeg})),
Eq. (\ref{xi2}) implies that

\beqa
\Omega & = & \frac{B}{R_-} ~~~.
\label{omega}
\eeqa

In order to find restrictions for the integration
constant $B$, we have to set some conditions,
namely:
i) Within the Schwarzschild radius, the expression (\ref{int-a})
has to be positive, such that the time is defined in the usual way
(no imaginary time, though in future one has to investigate the consequences
of an imaginary time, too). \\
ii) For large distances, the old equations of motion of
GR, ${\cal R}_{\mu\nu}$ = 0,
should arise, i.e., the equivalence to the standard variational principle
should emerge. This is because for large $r$ the Schwarzschild solution
should arise.

The condition ii) is automatically fulfilled: Using (\ref{xi1-xi2})
with (\ref{xi2}) and (\ref{omega}) leads for large $R_-$ to
$\xi_1 \rightarrow (2B)/R_-^4$. The function $\xi_0=\xi_1$ has the same form,
$\xi_3$ is proportional to $\xi_2$ and $\xi_2$ also vanishes for large $R_-$
like $1/R_-^2$ (see (\ref{xi2})). Thus, for $R_-$ very large, the $\xi$ functions tend to zero
and the results of standard GR are recovered.

The condition i) is satisfied for

\beqa
g^0_{00} & > & 0 ~~~.
\eeqa
As we will see in the next subsection, the $M_\pm$ values can
be set equal to $m$. In addition, taking only into account
terms up to $l^0$, the $R_\pm$ variables are both equal to the radial
distance $r$. Substituting this into
$g_{00}^0$ component (see (\ref{metric2}) below), we obtain a limiting value for $B$, solving
$g_{00}^0=0$, i.e.,

\beqa
g_{00}^0 & = & \left(1-\frac{2m}{r}+\frac{B}{2r^2} \right) ~=~ 0 ~~~.
\label{g00}
\eeqa
We are only allowing real solutions ($r_0$) for the radius variable $r$. 
The solution of the quadratic equation (\ref{g00}) is

\beqa
r_0 & = & m \left( 1 \pm \sqrt{1 - \frac{B}{2m^2}} \right) ~~~.
\label{sol}
\eeqa
When the square root in (\ref{sol}) is different from zero and positive,
then there are two real solutions $r_+$ and $r_-$,
where the index refers to the sign in (\ref{sol}).
Between these two real solutions the $g^0_{00}$ is negative,
which can easily be verified by substituting $r_0$ into (\ref{g00}).
A negative $g^0_{00}$ would break condition ii) above, thus, it
is excluded.
In order to avoid a negative $g^0_{00}$, the expression in the square
root can be at most 0, which implies $B=2m^2$ and, thus,
$r_0$ can have at most one real solution.
A negative value under the square root implies an imaginary $r_0$: For
this case, there is no real solution of the quadratic equation (\ref{g00})
and $g_{00}^0$ is always positive.
As we will see in section V, for $g^0_{00}$ = 0 the
redshift will be infinite at half of the Schwarzschild radius, implying
a physical division between the interior and exterior of this radius.
We require that all space is connected and, therefore, the limiting
solution of $g^0_{00}$ has to be excluded, too.
Nevertheless, we still will discuss this particular case in what follows,
interpreting it as a limit of $B=(2+\epsilon )m^2$, with a small
$\epsilon$.
Imposing the above restrictions leads to the condition

\beqa
B & > & 2m^2 ~~~.
\eeqa

For a further discussion, we need to enlist
the components of the average metric ($g^0_{\mu\nu}$) and,
for completeness, the
difference metric ($h_{\mu\nu}$). They are given by

\beqa
g^0_{00} & = & 1-\left[\frac{M_+}{R_+}+\frac{M_-}{R-}\right]+
\frac{\Omega}{2R_-} \nonumber \\
g^0_{rr}
& = & -\frac{1}{2} \left[ \frac{1}{\left( 1- \frac{2M_+}{R_+}\right)}
+ \frac{1}{\left( 1- \frac{2M_-}{R_-} + \frac{\Omega(R_-)}{R_-}\right)}
\right] \nonumber \\
g^0_{\theta\theta} & = & -\frac{1}{2}\left( R_+^2+R_-^2\right)
\nonumber \\
g^0_{\phi\phi} & = & -\frac{1}{2}\left( R_+^2+R_-^2\right)
{\rm sin}^2(\theta ) \nonumber \\
h_{00} & = & \frac{M_-}{R_-} - \frac{M_+}{R_+} - \frac{\Omega (R_-)}{2R_-}
\nonumber \\
h_{rr} & = & -\frac{1}{2}\left[ \frac{1}{\left(1- \frac{2M_+}{R_+}\right)}
- \frac{1}{\left(1- \frac{2M_-}{R_-}+\frac{\Omega(R_-)}{R_-}\right)}
\right] \nonumber \\
h_{\theta\theta} & = & -\frac{1}{2}\left(R_+^2-R_-^2\right) \nonumber \\
h_{\phi\phi} & = & -\frac{1}{2}\left(R_+^2-R_-^2\right)
sin^2(\theta ) ~~~.
\label{metric2}
\eeqa
All other elements of the metric are zero.

\subsection{Investigating an approximate solution for an exceptional case}

For simplicity, we will consider the limiting case of $B=2m ^2$,
stressing again that one has to add a small value in order that
$g^0_{00}$ is not exactly zero.
We analyze some consequences of the solution, obtained
above, applying it to the motion of a particle
at a distance of $r \approx m$ from the center,
where the $g^0_{00}$ becomes zero. For simplicity, we approximate
the length element $d\omega^2$ by the expression given in (\ref{domega}), i.e., only
the average metric $g^0_{\mu\nu}$ is considered and only
correction up to order $l^0$ are taken into account. This also
implies that $R_\pm \approx r$ = $r\pm l {\dot r}$.
In (\ref{domega}) the variables $\theta$ and $\phi$ reduce to real values,
with the usual interpretation as azimuthal and polar angles.

With this, the diagonal elements of the average metric,
as given in (\ref{metric2}), can be approximated by

\beqa
g^0_{00} & \approx & 1-\frac{M_++M_-}{r}+
\frac{B}{2r^2} \nonumber \\
g^0_{rr}
& \approx & - \frac{\left( 1-\frac{M_++M_-}{r} +\frac{B}{2r^2} \right)}
{(1-\frac{2M_+}{r})(1-\frac{2M_-}{r} +\frac{B}{r^2})}
\nonumber \\
g^0_{\theta\theta} & \approx & -r^2
\nonumber \\
g^0_{\phi\phi} & \approx & -r^2
{\rm sin}^2(\theta )                ~~~.
\eeqa
Taking into account, that for $B=0$ we should get back the standard
Schwarzschild metric, suggests the identification of the mass parameters
\cite{adler}

\beqa
M_+ & = & M_- ~=~ m ~~~,
\eeqa
with $m=\frac{GM}{c^2}$, with $G$ as the gravitational constant,
$M$ as the mass of the object and $c$ the light velocity.

We follow closely the steps as indicated in chapter 6.3 of \cite{adler}.
As shown in Eq. (\ref{domega}) the length element reduces to the usual
one ($d\omega^2 \approx ds^2 =g^0_{\mu\nu}dx^\mu dx^\nu$) with a modified real metric
$g_{\mu\nu}^0$. As a consequence, the steps to follow will be identical
to the standard description of GR. The contributions due to
the pseudo-complex structure are simulated by the appearance of the term
proportional to $B \ne 0$.

The variational procedure ($X^0 = ct$), which has to be applied now,
yields

\beqa
& \delta \int \left\{ (1-\frac{2m}{r} + \frac{B}{2r^2}) c^2{\dot t}^2
- \frac{ (1-\frac{2m}{r} + \frac{B}{2r^2})}
{(1-\frac{2m}{r})(1-\frac{2m}{r} + \frac{B}{r^2})} {\dot r}^2
\right. & \nonumber \\
& \left.
-r^2 ({\dot \theta}^2 + sin^2\theta~{\dot \phi}^2 ) \right\} ds  =  0 &
~~~.
\label{var2}
\eeqa
The dot indicates now a differentiation with respect to the curve parameter $s$.

Varying with respect to the variables $\theta$, $\phi$ and $t$, gives

\beqa
\frac{d}{ds} (r^2 {\dot \theta}) & = &
r^2 sin\theta ~ cos\theta ~{\dot \phi}^2 \nonumber \\
\frac{d}{ds}(r^2 sin^2\theta ~ {\dot \phi}) & = & 0 \nonumber \\
\frac{d}{ds}\left[ \left( 1-\frac{2m}{r} + \frac{B}{2r^2} \right)
{\dot t} \right]
 & = & 0 ~~~.
\label{eq-m}
\eeqa
The second equation gives the conservation of the angular momentum. This
implies that the motion is on a plane and one can choose
$\theta = \frac{\pi}{2}$. With this value,
the second equation in (\ref{eq-m}) yields $r^2 {\dot \phi} = h = const$.
The third equations yields

\beqa
\left( 1-\frac{2m}{r} + \frac{B}{2r^2} \right) {\dot t} & = & \gamma=const
\label{gamma}
~~~.
\eeqa

An additional condition is obtained by dividing the line element 
$ds^2$, as it appears in the integrand of (\ref{var2}), by itself,
in complete analogy to the subsection 6.2 of \cite{adler}, which gives

\beqa
1 & = & (1-\frac{2m}{r} + \frac{B}{2r^2}) c^2 {\dot t}^2
\nonumber \\
&&
 - \frac{\left( 1-\frac{2m}{r} + \frac{B}{2r^2} \right)}
{(1-\frac{2m}{r})(1-\frac{2m}{r} +\frac{B}{r^2})}
{\dot r}^2
\nonumber \\
&&
- r^2 ({\dot \theta}^2 + sin^2\theta ~ {\dot\phi}^2 ) ~~~.
\label{87}
\eeqa

Now, we denote a derivative with respect to the variable $\phi$ by
a prime. For example

\beqa
r^\prime & = & \frac{dr}{d\phi} ~=~ \frac{{\dot r}}{{\dot \phi}}
~~~.  \nonumber
\eeqa

From this equation and $r^2{\dot \phi}=h$, setting $\theta = \frac{\pi}{2}$
(motion in a plane), we obtain

\beqa
{\dot r} & = & {\dot \phi} r^\prime ~=~ \frac{h}{r^2} r^\prime
~~~.  \nonumber
\eeqa

Using (\ref{gamma}) and the above deduced relation of $r^2 {\dot \phi}=h$,
leads to

\beqa
1 & = & (1-\frac{2m}{r} + \frac{B}{2r^2})^{-1} c^2 \gamma^2
\nonumber \\
&&
 - \frac{\left( 1-\frac{2m}{r} + \frac{B}{2r^2} \right)}
{(1-\frac{2m}{r})(1-\frac{2m}{r} +\frac{B}{r^2})}
\frac{h^2}{r^4} r^{\prime~2} - \frac{h^2}{r^2}
~~~.  \nonumber
\label{x2}
\eeqa

As a next step the variable

\beqa
u & = & \frac{1}{r}  \nonumber
\eeqa
is introduced. The relation of the differential of $r$
with respect to $\phi$ to the one of $u$ is

\beqa
r^\prime & = & -\frac{u^\prime}{u^2}
~~~.                 \nonumber
\label{x3}
\eeqa
Multiplying (\ref{x2}) with $\frac{1}{h^2}(1-\frac{2m}{r})$
$\left( 1-\frac{2m}{r} + \frac{B}{r^2} \right)$ and using the
substitution (\ref{x3}), gives

\beqa
& \frac{1}{h^2}
\left( 1-2mu\right)\left( 1-2mu+Bu^2\right)  = & \nonumber \\
& \frac{\left(1-2mu\right)\left( 1-2mu+Bu^2\right)}
{h^2\left( 1-2mu+\frac{B}{2}u^2\right)}
c^2\gamma^2
- \left( 1-2mu+\frac{B}{2}u^2\right)u^{\prime~2} &
\nonumber \\
& -u^2\left(1-2mu\right)\left( 1-2mu+Bu^2\right) &
\nonumber
\eeqa
Dividing it by $\left( 1-2mu+\frac{B}{2}u^2\right)$ and solving for
$u^{\prime~2}$ leads to the equation of motion

\beqa
u^{\prime ~ 2} & = & \frac{c^2 \gamma^2}{h^2}
\frac{(1-2mu)(1-2mu+Bu^2)}{(1-2mu+\frac{B}{2}u^2)^2} \nonumber \\
&& -u^2 \frac{(1-2mu)(1-2mu+Bu^2)}{(1-2mu+\frac{B}{2}u^2)} \nonumber \\
&& -\frac{(1-2mu)(1-2mu+Bu^2)}{(1-2mu+\frac{B}{2}u^2) h^2}
~~~.
\label{up2}
\eeqa
Setting $B=0$ results in the standard equation of GR for
the planetary motion.

This equation is still
difficult to solve. We, therefore, recur to
a further approximation. In the last
section we showed that for $B=2m^2$ at $r=m$
the metric component $g^0_{00}$ becomes zero.
{\it This is finally the exceptional case we will study, expanding $u$ around
this minimum}, i.e.,

\beqa
u & = & \frac{1}{m} \left( 1 + \frac{\varepsilon}{2} \right) ~~.
\eeqa
The factor $\frac{1}{2}$ in front of $\varepsilon$ is for convenience.

We obtain for some important factors, appearing
in (\ref{up2}), setting $B=2m^2$,

\beqa
(1-2mu) & = & -(1+\epsilon ) \nonumber \\
(1-2mu+\frac{B}{2}u^2 ) & = & \frac{\epsilon^2}{4} ~~~.
\eeqa

This helps to determine the constant $\gamma$, introduced in
(\ref{gamma}). Inspecting the equation (\ref{gamma})
in (\ref{eq-m}) leads to

\beqa
\frac{\epsilon^2}{4} {\dot t} & = & \gamma ~~~.
\eeqa
This has to be fulfilled for any value of $\epsilon$, especially when
$\epsilon = 0$. Therefore, for $r$ near $m$ and
$B=2m^2$ the $\gamma$ has to be zero! This is valid in particular for
orbitals around $r=m$.

Taking into account only the leading terms in (\ref{up2}), we arrive at

\beqa
(\epsilon^\prime )^2 & \approx & \frac{16 m^2}{\epsilon^2}
\left[ \frac{1}{h^2} + \frac{1}{m^2} \right] ~~~.
\eeqa
Deriving it again, yields

\beqa
2\epsilon^\prime \epsilon^{\prime\prime} & \approx &
-\frac{32 m^2}{\epsilon^3} 
\left[ \frac{1}{h^2} + \frac{1}{m^2} \right] \epsilon^\prime ~~~.
\eeqa
There are two solutions. The first corresponds to $\epsilon^\prime =0$, i.e
$\epsilon = const$. This is the circular motion around the heavy mass
object at a distance $r=m$.
The other one is obtained for $\epsilon^\prime \neq 0$. We get

\beqa
\frac{d^2\varepsilon}{d\phi^2} & \approx &
-\frac{16 m^2}{\varepsilon^3}
\left[ \frac{1}{h^2} + \frac{1}{m^2} \right] ~~~.
~~~.
\label{eq-e}
\eeqa
Without solving it, we can already deduce
some properties from this equation.
When $\epsilon < 0$ ($r>m$) the acceleration $\epsilon^{\prime\prime}$
is positive, which translates into $r^{\prime\prime}$ negative, i.e., the
particle is accelerated toward the center. In contrast, when $\epsilon > 0$
($r<m$) the acceleration $\epsilon^{\prime\prime}$ is negative, which
translates into $r^{\prime\prime}$ {\it positive}, i.e., the particle
is {\it accelerated away from the center}.

We emphasize: {\it At $r>m$ a particle experiences
an attraction toward the center,
while at $r<m$ there is a repulsion!} This implies an \textbf{anti-gravitational interaction} for $r<m$!
In different words: a massive body continues to contract to lower values
of the Schwarzschild radius, until for $B=2m^2$ it reaches $r=m$.
For smaller values the heavy mass object feels repulsion.
The body may realize an oscillatory type of motion around $r=m$, but this
motion is not easy to describe, as can be seen even by the above
simplified equation of
motion.

The details change when $B>2m^2$, but the cross structure remains.

The contribution of $B$ seems to be equivalent to the introduction
of a $r$-dependent cosmological function $\Lambda$ (r), though, here it has a different
origin and details have still to be worked out. It would be interesting
to develop a model for the evolution of the universe, resolving the
modified Einstein equations with the contributions of $B$ and assuming a constant
mass distribution in the universe (Robertson-Walker).
What will be the possible dependences of
$\Lambda$ as a function in time and what will be its value? This
consideration we will leave for a later publication.

\section{The redshift}

In this section we calculate the deviation of the redshift,
comparing the present theory with GR. The redshift is an important observable and the
detection of possible deviations to known results
might be in reach for experiment in near future.

Of particular interest for the redshift is $g^0_{00}$. For distances
larger than the Schwarzschild radius, the solution is very similar to the
standard Schwarzschild solution. However, differences will appear
near and below the Schwarzschild radius. {\it First of all, there is no
singularity!} In addition, inside this radius the time component
of the metric is positive definite.

Taking all spatial distances to zero ($dr=d\theta = d\phi =0$), we have
to lowest order in $l$ that
$d\omega^2 \approx d\tau^2 \approx g^0_{00}(r) dt^2$,
with $\tau$ as the eigen-time.
From this we obtain for the change of frequency

\beqa
\nu & \approx & \sqrt{g^0_{00}(r)} ~ \nu_0 ~~~,
\eeqa
where $\nu_0$ is the frequency of a photon at the emission point
$r$ and $\nu$ is the observed one at large distance.

The redshift $z$ is defined as \cite{adler,pleba}

\beqa
z & = \frac{\nu_0}{\nu} - 1 ~=~ \frac{1-\sqrt{g^0_{00}}}{\sqrt{g^0_{00}}}
~~~.
\eeqa

In Ref. \cite{muller-wold} the $g$-factor is defined, which is in the
following relation to the redshift:

\beqa
g & = & \frac{\nu}{\nu_0} ~=~ \frac{1}{1+z} ~~~.
\eeqa
$g$ = 1 corresponds to a flat space ($z=0$), while $g < 1$ indicates the
relativistic effect ($z>0$) and $g=0$ corresponds to an infinite redshift
($z=\infty$).

\begin{figure}[ph]
\centerline{\epsfxsize=10cm\epsffile{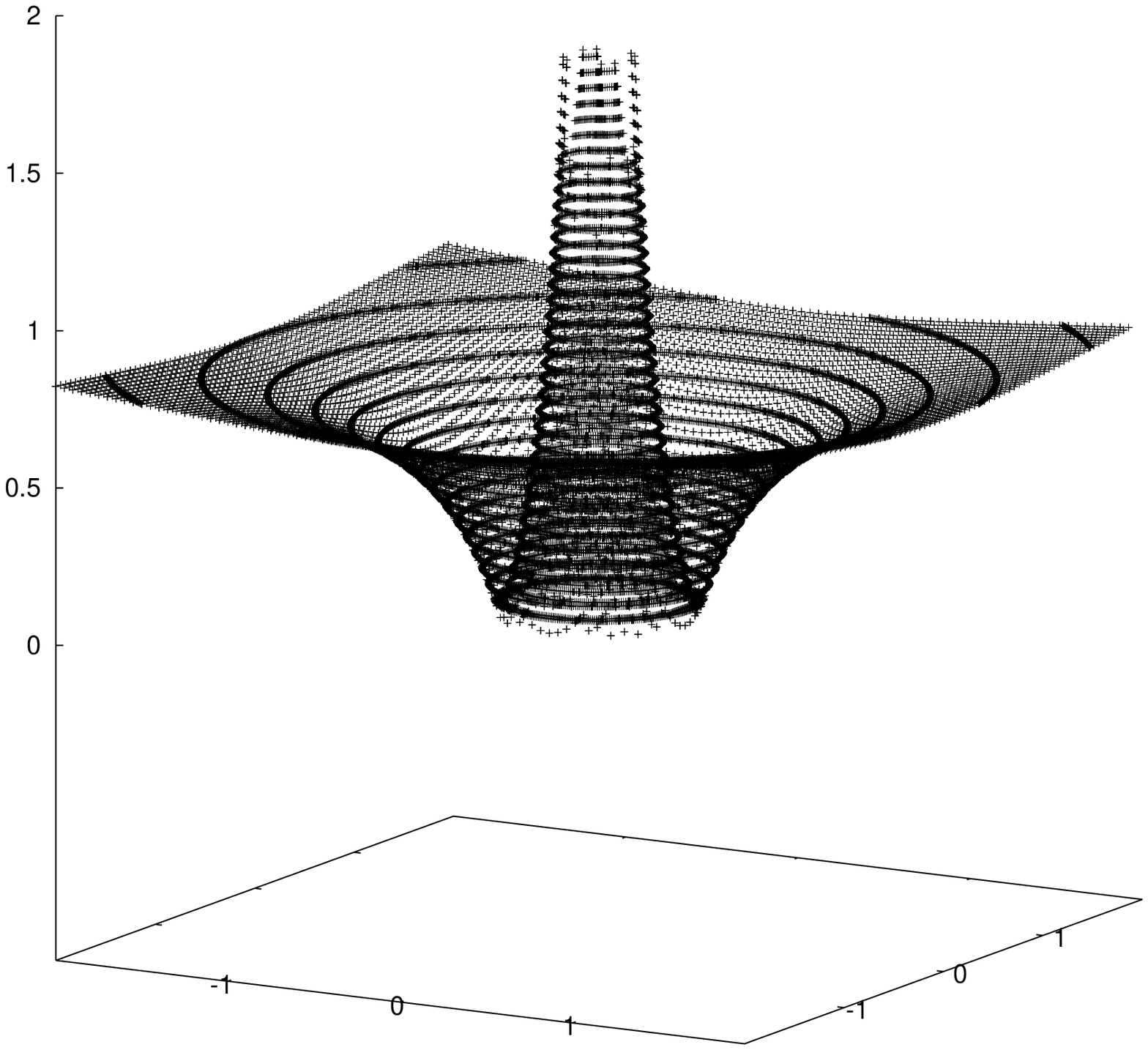}}
\caption{
The $g$-factor $g=\frac{\nu}{\nu_0}$, for $B=2 m^2$,
as a function of the coordinates
$x$ and $y$, in units of the Schwarzschild radius $2m$ ($\frac{r}{2m}=1$),
where $r$ is the radial distance given by $r$ = $\sqrt{x^2 + y^2}$,
The distance between two contours corresponds to a step size of 0.05.
}
\label{fig1}
\end{figure}

\begin{figure}[ph]
\centerline{\epsfxsize=10cm\epsffile{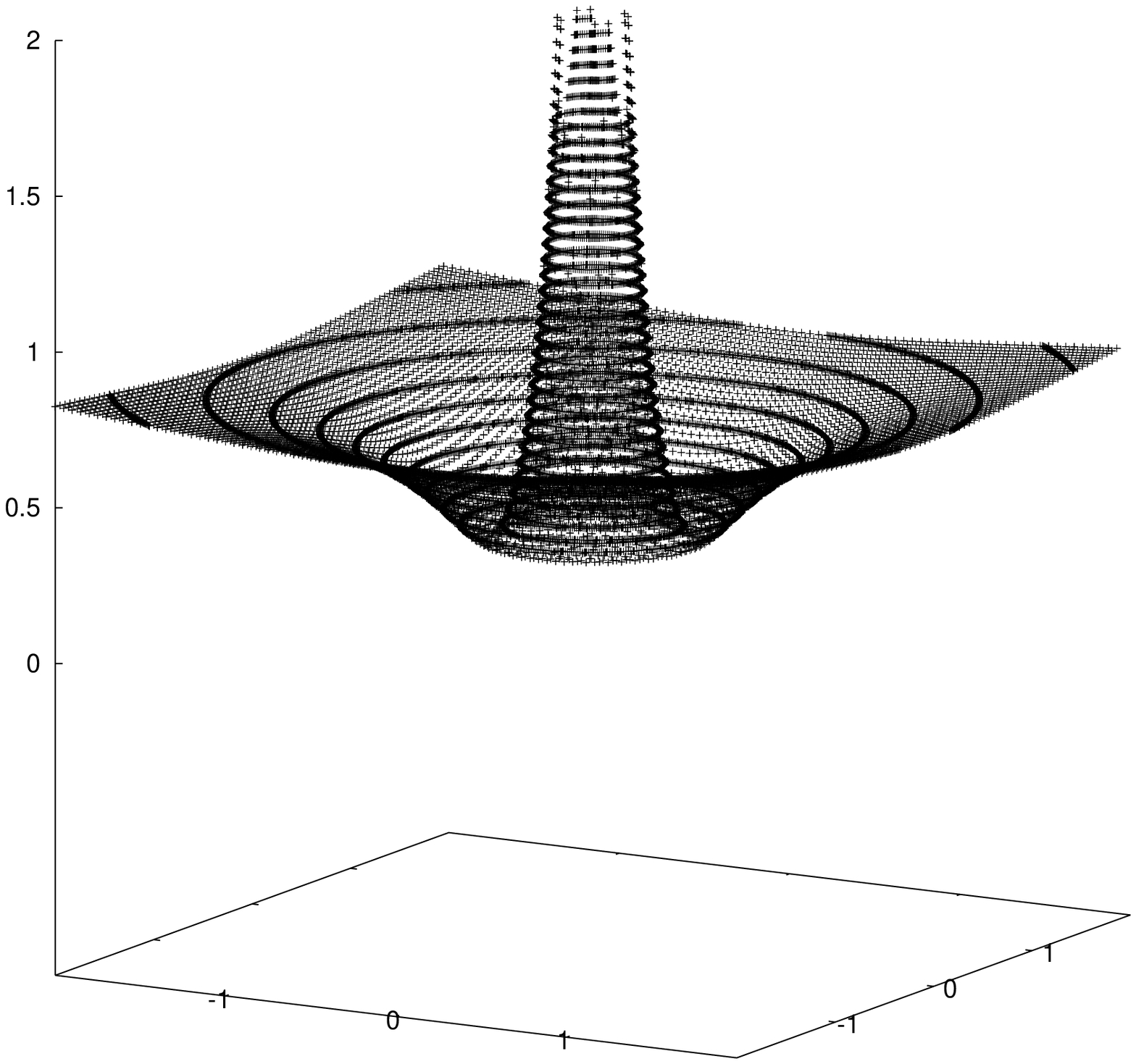}}
\caption{
The $g$-factor $g=\frac{\nu}{\nu_0}$, for $B=2.2 m^2$,
as a function of the coordinates
$x$ and $y$, in units of the Schwarzschild radius $2m$ ($\frac{r}{2m}=1$),
where $r$ is the radial distance given by $r$ = $\sqrt{x^2 + y^2}$,
The distance between two contours corresponds to a step size of 0.05.
}
\label{fig2}
\end{figure}

\begin{figure}[ph]
\centerline{\epsfxsize=10cm\epsffile{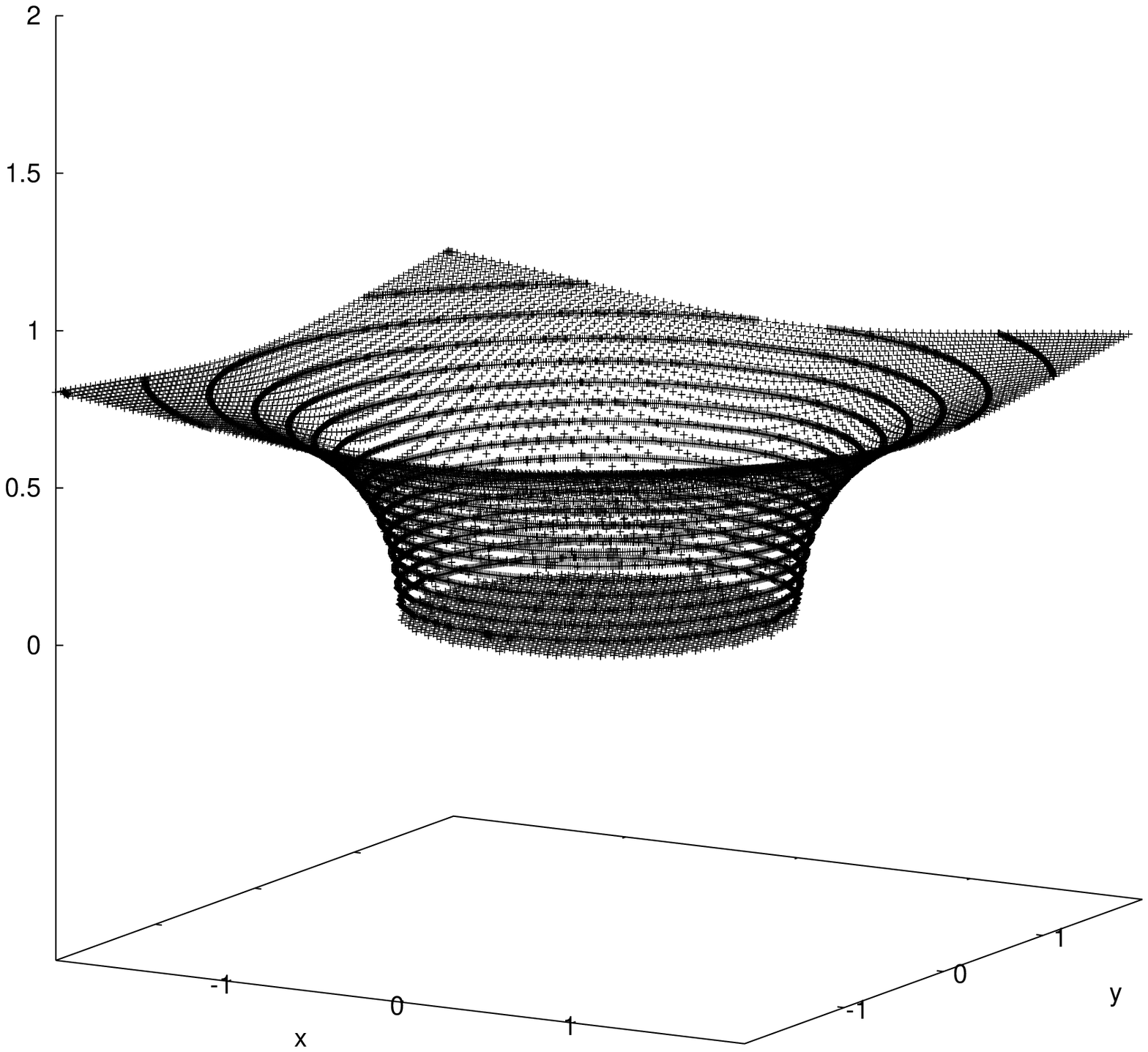}}
\caption{
The $g$-factor $g=\frac{\nu}{\nu_0}$, for $B=0$ the Schwarzschild
solution in the standard theory,
as a function of the coordinates
$x$ and $y$, in units of the Schwarzschild radius $2m$ ($\frac{r}{2m}=1$),
where $r$ is the radial distance given by $r$ = $\sqrt{x^2 + y^2}$,
The distance between two contours corresponds to a step size of 0.05.
The Schwarzschild solution is only valid up to $r=2m$ where the $g$-factor
acquires the value 0.
}
\label{fig3}
\end{figure}

Let us discuss the consequences for the solution obtained in the former
section: We obtain

\beqa
\nu & \approx & \sqrt{1-\frac{2m}{r}+\frac{B}{2r^2}}~~~\nu_0  ~~~.
\eeqa
The $g$-factor is just the square root expression in front of $\nu_0$.
For the two cases of $B=2m^2$ and $B=2.2m^2$,
its behavior is depicted in Figs. \ref{fig1}, \ref{fig2} respectively.
Please, note the strong anti-gravitational behavior below half
of the Schwarzschild radius.
For comparison, the $g$-factor for the Schwarzschild solution is
depicted in Fig. \ref{fig3}.
For $B=2.2m^2$ there is no real solution of
$g_{00}^0=0$, i.e., $g^0_{00}$ is always positive.
Using the value $B=2m^2$, produces a zero at $x=\frac{r}{2m}=0.5$, i.e.,
an infinite
redshift. The value of $B$ cannot be smaller, otherwise $g^0_{00}$ would
become negative for a certain the range of $r$.
Larger values of $B$ will produce $g$-factors which
are always larger than zero, with no infinite redshift.

For $r$ toward 0 a blueshift is obtained.
The blueshift is an effect of the anti-gravitational force as deduced
in the last section.

\begin{table}
\begin{center}
\begin{tabular}{|c|c|c|c|}
\hline
$\frac{r}{2m}$ & Schwarzschild & $B=2m^2$ & $B=2.2m^2$   \nonumber \\
\hline
0.125 & - (-) & 3.00 (-0.67) & 3.26 (-0.69) \\
0.25 &  - (-) & 1.00 (0.00) & 1.18 (-0.15)  \\
0.50 &  - (-) & 0.00 ($\infty$) & 0.32 (2.13) \\
0.75 &  - (-) & 0.33 (2.00) & 0.39 (1.56) \\
1.00 & 0.00 ($\infty$) & 0.5 (1.) & 0.52 (0.92)  \\
1.25 & 0.45 (1.22) & 0.60 (0.67) & 0.61 (0.64) \\
1.50 & 0.58 (0.72) & 0.67 (0.50) & 0.67 (0.49) \\
1.75 & 0.65 (0.54) & 0.71 (0.40) & 0.72 (0.39) \\
2 & 0.71 (0.41) & 0.75 (0.33) & 0.75 (0.33) \\
3 & 0.82 (0.22) & 0.83 (0.20) & 0.83 (0.20) \\
4 & 0.87 (0.15) & 0.88 (0.14) & 0.88 (0.14) \\
5 & 0.89 (0.12) & 0.90 (0.11) & 0.90 (0.11) \\
\hline
\end{tabular}
\end{center}
\caption{
Some key values of the $g$-factor and the redshift (in parenthesis)
as a function in 
$x=\frac{r}{2m}$, using $B=2m^2$ and $B=2.2m^2$.
For $r<2m$ ($x<1$) the standard Schwarzschild solution
does not exist and, therefore, it is indicated by "-".
}
\label{table1}
\end{table}

In Table I, some key values, like the $g$-factor and redshift (in parenthesis)
are given for several values of $r$.
From this table we also see, that at the distance $r=4m$ from the center,
the deviation from the standard Schwarzschild solution is still minimal.

Note, that the redshift is finite and should be measurable near the
Schwarzschild radius of giant masses in the center of
active galaxies.  In \cite{muller-wold} a method is presented how to
deduce from broad X-ray emission lines the redshift as a function
of the radial distance. Results of measured redshifts for the galaxy
MrK110 are presented. Unfortunately, the closest distance reported is
twice the Schwarzschild radius. Inspecting Table I shows that a
notable difference between our calculations and the standard
Schwarzschild solution only appears below this distance.

In the extended GR no singularities appear and, thus,
black holes in the literary sense do not exist.
Very large mass concentrations should appear instead and will be
rather gray.

\section{Conclusions}

We have presented a possible algebraic extension for the theory of General
Relativity, which does not contain singularities.
Pseudo-complex variables and a modified variational principle were
used.
We obtained a solution, which depends on the additional parameter $B$,
whose origin is in the pseudo-complex description.
We cannot determine the exact values of $B$, because it has to be
measured experimentally by detecting deviations of, e.g., the redshift
as obtained in our theory with respect to standard GR.

A first finding is the deviation of the
redshift compared to standard GR.
The calculated redshifts are not infinite any more but
approach finite values near the Schwarzschild radius.
The differences to the standard Schwarzschild solution are small
up to $r \approx 4m$. With the present state of technology
\cite{muller-wold}, however,
there is a good chance that the deviations may be observed
in near future.
A measured deviation from the standard solution
implies that large central masses appear as rather gray objects.

As a second important result we obtained an anti-gravitational effect for
radii smaller than half of the Schwarzschild radius, assuming
the particular values $B=2m^2$ and $B=2.2m^2$. For other values of $B$ the
scenario is similar, i.e., for $r$ smaller than a given distance,
anti-gravitation appears. As a consequence, heavy mass objects can not
contract to a point at $r=0$. The origin of this
effect will be further discussed in a forthcoming paper.

The formulation of the extended GR is done in complete
analogy to standard GR, which is advantageous.
The difference to standard GR is the use of
two, in general distinct, metrics $g^\pm_{\mu\nu}$.

We showed that the algebraic extension of GR to pseudo-complex variables
can be achieved in a consistent manner. The formulation permits non-singular
solutions. A unique solution for a spherically
symmetric mass distribution was obtained, with the consequences discussed
above. Whether the theory is realized in nature remains
to be verified by
experiment. Possible signatures in the redshift are proposed.

\section*{Acknowledgement}
P.O.H. would like to thank the
{\it Frankfurt Institute for Advanced Studies} (FIAS)
at Frankfurt am Main for the hospitality and in particular for the
excellent atmosphere during his stay in June 2008.
We acknowledge financial support from DGAPA-UNAM and CONACyT.
Useful discussions with A. Bounames and K. Nouicer are acknowledged.

\end{document}